# Independent and simultaneous tailoring of amplitude, phase, and complete polarization of vector beams

**Hao Chen[1], Tianxiang Huang[2], Jianping Ding[3] , and Guoqiang Li[1, 2, *]**

*[1]Department of Ophthalmology and Visual Science, The Ohio State Universit,y, Columbus, OH 43212, USA*

*[2]Department of Electrical and Computer Engineering, The Ohio State University, Columbus, OH 43212, USA*

*[3]National Laboratory of Solid State Microstructures and School of physics, Nanjing University, Nanjing 210093, China*

*[*] Corresponding author: li.3090@osu.edu*

*(This work was done at OSU in 2012)*

We present an approach that enables complete control over the amplitude, phase and arbitrary polarization state on the Poincaré sphere of an optical beam in a 4-$f$ system with a spatial light modulator (SLM). The beams can be constructed from a coaxial superposition of x- and y-linearly polarized light, each carrying structured amplitude profile and phase distributions by using an amplitude-modulated mask imposed on the SLM. The amplitude, phase and polarization distribution of vector beams with four free parameters can be tailored independently and simultaneously by the SLM.

**Key words:** Physical optics; Polarization; Phase modulation; Spatial light modulators; Holographic interferometry.

Control of the optical beam with spatially variant polarization and phase distribution has become a rapid growing topic of interest in recent years, owing to the quest for higher resolution microscopy and potential applications in a variety of realms such as polarimetry, microlithography, focus engineering and optical tweezers [1-7]. Besides the polarization and phase distributions, amplitude distribution of the optical field is another important parameter, since some focal field distributions cannot be attained with polarization control alone. It is found that arbitrary 3D state of polarization control with a diffraction-limited spot size can be obtained through modified amplitude and polarization distribution of an incident beam [8]. Motivated by these applications, various methods for generating vector beams have been proposed [7, 9-16]. However, most of these reported methods have limitations owing to the fact that no more than three parameters of the optical field can be controlled. Therefore, the polarization states are limited to a line on the Poincaré sphere (PS) or yielded to the amplitude distribution of the input beam due to the loss of one freedom of polarization − phase or amplitude. Complete control of an optical field [17] requires modulation of four independent parameters, one for amplitude, one for phase retardation and two for polarization mapping onto the surface of Poincaré sphere. In this Letter, we propose a novel approach for generating vector beams with the desired amplitude, polarization and phase structures with four free parameters. Our scheme enables space-variant modulation of the light beams in amplitude, phase and the full polarization ellipse, and is very helpful in expanding the function of the vector beams.

The general experimental arrangement is shown in Fig. 1. A collimated linearly polarized light is incident onto the spatial light modulator (SLM) which is used to encode a holographic grating (HG) pattern. The diffracted light enters a 4*f* system composed of a pair of identical lenses with a focal length of *f*. The SLM is located at the front focal plane of the first lens. The two-dimensional HG is displayed on the SLM and the incident light is diffracted into various orders. The grid-like transmission function of the SLM is written as

$$t\left(x,y\right)=\left\{\frac{1}{2}+\frac{1}{4}\left[A_1\left(x,y\right)\cos\left(2\pi f_0\,x+\phi_1\left(x,y\right)\right)+A_2\left(x,y\right)\cos\left(2\pi f_0\,y+\phi_2\left(x,y\right)\right)\right]\right\}, \qquad (1)$$

where $\phi_1\left(x,y\right)$ and $\phi_2\left(x,y\right)$ are the additional phase distribution imposed on the vertical and horizontal HGs respectively, and $f_0$ is the spatial frequency of the HGs. $A_1\left(x,y\right)$ and $A_2\left(x,y\right)$ are the local modulation depths of the vertical and horizontal HGs respectively and are defined as $0\le A_1\left(x,y\right)\le 1, 0\le A_2\left(x,y\right)\le 1$. As the modulation depth distributions $A_1\left(x,y\right)$ and $A_2\left(x,y\right)$ are changed, the diffraction efficiency of the filter is modified and this provides additional amplitude modulations of the two first-order diffractions in the orthogonal directions in the Fourier plane. Light that is not diffracted into the first order is sent to the zero order, effectively allowing for amplitude modulation of both first-order diffraction efficiencies. Only the first orders are allowed to pass through a spatial filter (with two separate open apertures) placed at the Fourier plane of the 4-*f* system, and these two terms are $\mathbf{F}\left\{A_1\left(x,y\right)e^{i\phi_1(x,y)}\right\}\otimes\delta\left(f_x-f_0,f_y\right)$ and $\mathbf{F}\left\{A_2\left(x,y\right)e^{i\phi_2(x,y)}\right\}\otimes\delta\left(f_x,f_y-f_0\right)$, where $\mathbf{F}\left\{\cdots\right\}$ represents the two-dimensional Fourier transform and $\otimes$ represents the convolution operation. Then they are converted by two λ/2 wave plates into the x- and y linearly polarized beams, respectively. After being Fourier transformed by the second converging lens, the two orders turn to $A_1\left(x,y\right)e^{i\phi_1(x,y)}\left(1,0\right)^T, A_2\left(x,y\right)e^{i\phi_2(x,y)}\left(0,1\right)^T$, and collinearly recombined at the rear plane of the 4f system by a Ronchi grating. The output beam $\vec{E}$ can be expressed by the superposition of x- and y-linear polarization components as

$$\vec{E}(x,y)=\begin{pmatrix}E_x(x,y)\\E_y(x,y)\end{pmatrix}=\left[e^{i\phi_1(x,y)}\begin{pmatrix}A_1(x,y)\\0\end{pmatrix}+e^{i\phi_2(x,y)}\begin{pmatrix}0\\A_2(x,y)\end{pmatrix}\right]=A(x,y)e^{i\beta(x,y)}\begin{pmatrix}\cos\alpha(x,y)\\\sin\alpha(x,y)e^{i\gamma(x,y)}\end{pmatrix},\qquad(2)$$

where $A=\sqrt{A_1^2+A_2^2}$ , $\cos\alpha=\dfrac{A_1}{A}$ , $\beta=\phi_1$, and $\gamma=\phi_2-\phi_1$. The normalized amplitude, phase and state of polarization (SOP) of the vector beam can be tailored by $A(x,y)$ ,

$\beta(x,y),\alpha(x,y)$ and $\gamma(x,y)$, respectively. Two parameters $\alpha(x,y)$ and $\gamma(x,y)$ are used

to define the point on the surface of PS. The PS is constructed with $S_0$ as the unit radius from

the origin and using $S_1$, $S_2$, $S_3$ as the sphere's Cartesian coordinates. The sphere's spherical

angles ($2\psi$ , $2\chi$ ) are given by [18]

$$\psi=\frac{1}{2}\tan^{-1}\left(\tan2\alpha\cos\gamma\right),\chi=\frac{1}{2}\sin^{-1}\left(\sin2\alpha\sin\gamma\right).\qquad(3)$$

$2\psi$ and $2\chi$ stand for the latitude and longitude angles of this point in the spherical coordinate

system, respectively (Fig 2 (a)). It is easily observed from Eq. (2) that if one prescribes the parameters $\alpha(x,y)$ and $\gamma(x,y)$, a beam with arbitrary spatially variant polarization and all possible states of polarization can be obtained.

Equation (2) is the fundamental principle of this paper, which suggests that four free parameters A, $\alpha$, $\beta$ and $\gamma$ could be used for independent modulation of the amplitude,

polarization and phase of the vector beams. To realize the full control of the vector fields

(consisting of linear, circular and elliptical polarizations), the experimental arrangement is

very similar to Fig. 1 in Ref. 19. Two λ /2 wave plates at the back focal plane of the first

Fourier transform lens are used to transfer the two first orders diffracted by SLM into the x-and y- linearly polarized fields as a pair of orthogonal base vectors, each carrying different amplitude and phase information.

To validate the feasibility of the proposed approach, we present a few examples to demonstrate our capability of manipulating the amplitude, phase and complete polarization structure simultaneously. We begin with a special case of the triple concentric mode vector beams. The equally separated modes with uniform amplitude distribution correspond to

$$\vec{E}_1(x,y) = \begin{pmatrix} E_x(x,y) \\ E_y(x,y) \end{pmatrix} = \begin{pmatrix} e^{i\phi(x,y)} \\ e^{-i\phi(x,y)} \end{pmatrix}, \quad \vec{E}_2(x,y) = \begin{pmatrix} E_x(x,y) \\ E_y(x,y) \end{pmatrix} = \begin{pmatrix} \cos\phi \\ i\sin\phi \end{pmatrix}, \quad \text{and} \quad \vec{E}_3(x,y) = \begin{pmatrix} E_x(x,y) \\ E_y(x,y) \end{pmatrix} = \begin{pmatrix} \cos\phi \\ \sin\phi \end{pmatrix} \quad \text{from}$$

inner mode to the outer mode. The three modes ($\vec{E}_1, \vec{E}_2, \vec{E}_3$) in this example correspond to the circular lines of two orthogonal meridian and equator, each of which rotates around the perpendicular axes of the PS (Fig 2.a). To characterize the distribution of SoPs of a vector field, four Stokes parameters $S_0$, $S_1$, $S_2$ and $S_3$, in the representation of the PS should be specified. The measured results, as shown in Fig 2 (c) through (f), exhibit fan-like pattern, implying that SoPs are indeed azimuthally variant as predicted. The three orthogonal circular lines on the PS demonstrate that all states of polarization can be produced by our method.

In order to get insights into the phase structure, a Ronchi amplitude grating is placed in front of the CCD camera so as to create a self interference pattern on the CCD camera. The generated vector field illuminates the amplitude grating which diffracts the incoming light into various diffraction orders. When the frequency of the grating is low, interference pattern is generated by the slightly detached diffraction orders interfering with each other [19]. The interference pattern with straight fringes corresponds to the vector beam with plane wavefront for comparison with the example below. In doing so, we do not directly observe the phase of the vector beam itself but rather indirectly confirm its existence through the self interference fringes behind the amplitude grating resulting from superposition of various diffraction orders.

Second, we investigate a vector beam with its cross-sectional field described as $\vec{E}_1(x,y) = \begin{pmatrix} \cos\phi \\ \sin\phi \end{pmatrix} e^{\pm i 2\phi}$, $\vec{E}_2(x,y) = 0.8 \begin{pmatrix} e^{i\phi(x,y)} \\ e^{-i\phi(x,y)} \end{pmatrix}$, and $\vec{E}_3(x,y) = 0.6 \begin{pmatrix} \cos\phi \\ i\sin\phi \end{pmatrix}$. The amplitude of each equally separated mode is step declined, and the ratio is 1:0.8:0.6 as illustrated in the Fig. 3 (a). This example shares the same polarization mode as the first one but with a switch of the positions. Moreover, an additional phase distribution is imposed on the inner mode. As illustrated above, the interference patterns captured by the CCD camera are staggered in the center. The phase structure of the generated vector beam with helical charge (±2) corresponds to the shifted fork fringe in the opposite direction (Fig. 3 (f) – (i)).

Figure 4 shows the last example for generation of the vector beams with Bessel amplitude envelope, which is one of the most studied optical field distributions [20]. Two types of radially polarized first-order Bessel beams are expressed by $\vec{E}_1(x,y) = \left| J_1(\rho,\varphi) \right| (\cos\varphi, \sin\varphi)^T$, $\vec{E}_2(x,y) = \left| J_1(\rho,\varphi) \right| (\cos\varphi, \sin\varphi)^T \exp(i2\varphi)$, respectively, where $J_p(\cdot)$ denotes the Bessel function of the first kind with order p, $r_0$ is the radius of the aperture-truncated beam. The two types of Bessel beams share the same amplitude profile and polarization state but have different phase structures. In Fig. 4(a), we show the intensity profiles of the two cases in the radial direction and the theoretical Bessel profile is also shown for comparison. The straight and fork fringes in the interference patterns correspond to the plane wavefront and helical wavefront of the two vector beams, respectively.

To summarize, we have proposed a flexible and powerful method for generating vector beams with which we can achieve the desired amplitude, phase and all possible polarization states on the PS. A significant advance made in this work is that the amplitude, phase and SoP of the vector beams can be modulated arbitrarily, independently and dynamically, by using only one SLM. With the complete control parameters of our setup, it is more flexible to manipulate the optical field and thus it greatly helps us to further explore the engineering of the vector beams for specific applications, such as vector point spread function design with extended focal field and functional optical tweezers.

**Figure captions**

Fig. 1. (Color online) Schematic of the experimental setup.

Fig. 2. (Color online) Generated vector beam with triple concentric modes. (a) Illustration of three rotation trajectories of the polarization states around the three perpendicular axes of the Poincaré sphere. The polarization rotations of the inner mode, middle mode and outer mode correspond to blue, green and yellow circular traces on the Poincaré sphere. (b) Space-variant SoP. (c-f) Stokes parameters $S_0$, $S_1$, $S_2$ and $S_3$. (g) Prescribed phase profile. (h) Interference pattern immediately after the amplitude Ronchi grating.

Fig. 3. (Color online) Generated vector beam with step declined amplitude. (a) Space-variant SoP. (b-e) Stokes parameters $S_0$, $S_1$, $S_2$ and $S_3$. (f), (h) The prescribed phase profiles with helical charge ($\pm 2$). (g), (i) The corresponding interference patterns immediately after the amplitude Ronchi grating.

Fig. 4. (Color online) Generated Bessel vector beams. (a) Amplitude profile in the radial direction. (b-f), (j-n) CCD camera images of both vector fields with or without passing through a linear polarizer. The orientation of the analyzer is indicated in the inset of each panel. (g, o) are the prescribed phase profiles of the plane wavefront and helical wavefront respectively. (h, i), (p, q) are the simulated and experimental results of the interference pattern immediately after the amplitude Ronchi grating.

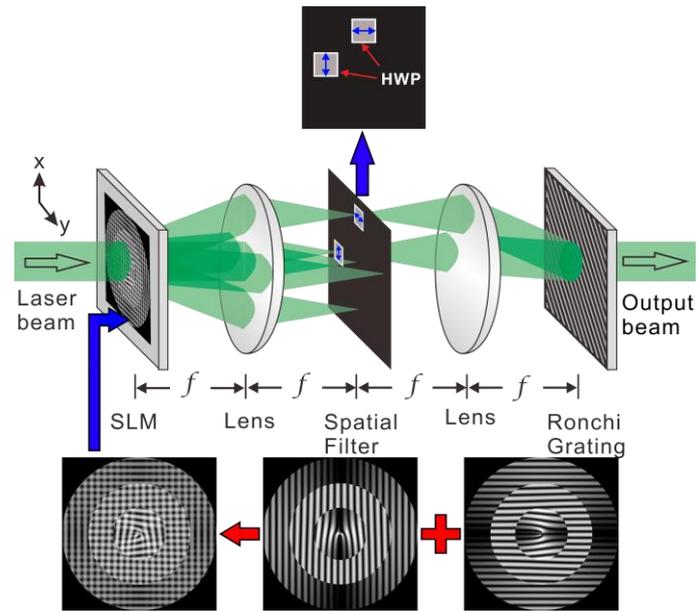

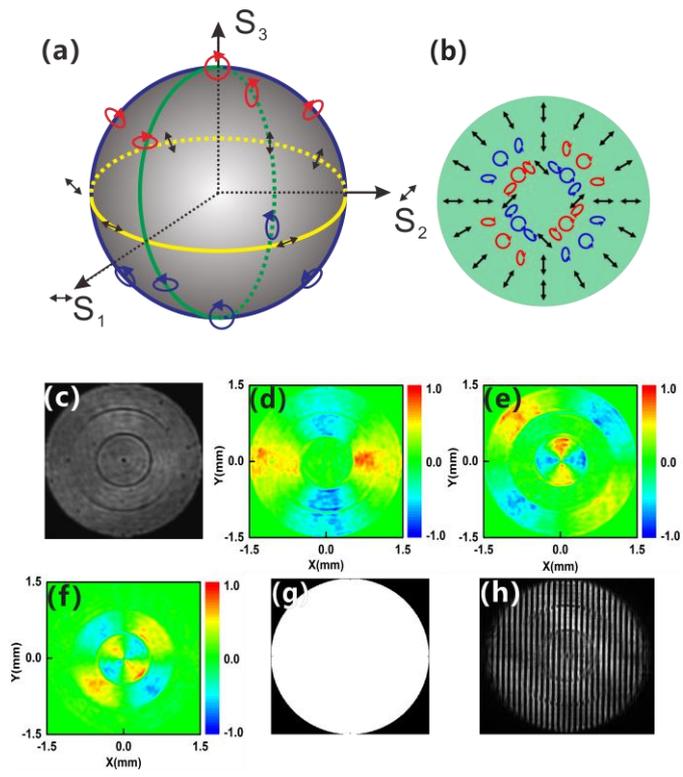

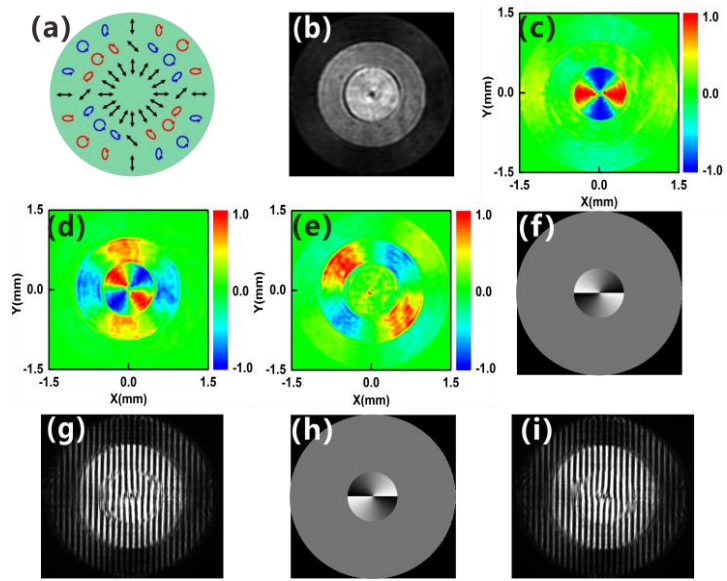

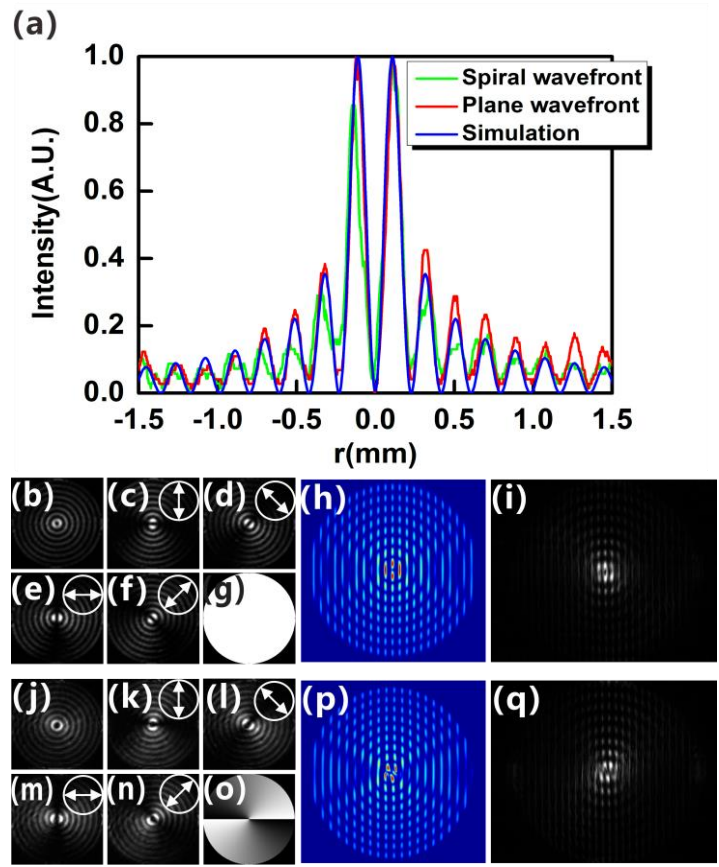